# Relativistic transformation of temperature and Mosengeil-Ott's antinomy

J. J. Mareš[1], P. Hubík, J. Šesták, V. Špička, J. Krištofik, and J. Stávek

Institute of Physics of the ASCR, v. v. i., Cukrovarnická 10,
162 00, Prague 6, Czech Republic

**Abstract**

A not satisfactorily solved problem of relativistic transformation of temperature playing the decisive role in relativistic thermal physics and cosmology is reopened. It is shown that the origin of the so called Mosengeil-Ott's antinomy and other aligned paradoxes are related to the wrong understanding of physical meaning of temperature and application of Planck's Ansatz of Lorentz's invariance of entropy. In the contribution we have thus reintroduced and anew analyzed fundamental concepts of hotness manifold, fixed thermometric points and temperature. Finally, on the basis of phenomenological arguments the Lorentz invariance of temperature and relativistic transformations of entropy are established.

## 1. Introduction

There is a non trivial problem connected with the relativistic transformation of temperature which can be, however, reduced to a simple question: Is the body moving with the velocity $v$ relatively to the rest system of coordinates in its own coordinate system colder or hotter than if it were measured in the rest system of coordinates? Very early after the appearance of the special theory of relativity in 1905 the problem was solved by a pupil of M. von Laue, K. von Mosengeil, who provided the following result [1]:

$$T = T_0 \sqrt{(1 - v^2/c^2)}, \quad \text{(Mosengeil, 1907)} \qquad (1)$$

where $T_0$ is Kelvin's temperature as measured in the rest system of coordinates and $T$ the corresponding temperature detected in the moving system. The satisfaction with this formula which is up to now serving as a standard in textbooks on special theory of relativity was put in doubts by a challenging paper due to H. Ott [2] where just an inverse formula for the relativistic transformation of temperature is derived, namely

$$T = T_0 / \sqrt{(1 - v^2/c^2)}. \quad \text{(Ott, 1963)} \qquad (2)$$

---

[1] maresjj@fzu.cz

After the appearance of Ott's paper a vivid discussion in "Nature" [3-8] broke out which, however, stopped after some time without bringing any clear decision which of these two formulae is true [9, 10]. At the end of the 20$^{th}$ century, during the last wave of interest in the problem [11, 12], however, another opinion appeared, namely that the temperature must be Lorentz invariant [13].

As we believe for the solution of this fundamental problem of relativistic thermal physics, it is inevitable to make first clear what the temperature actually is and what it is not. Therefore, before suggesting our solution of this interesting puzzle, we are critically revising the definition of fundamental concept of phenomenological thermal physics, i.e. that of the temperature.

## 2. The concept of temperature

Astonishingly, a satisfying definition of the phenomenological physical quantity called temperature is lacking in the literature. There exist, of course, definitions in terms of statistical physics, which are, however, related to the phenomenological quantity measured by a macroscopic device, thermometer, only by rather superficial considerations. Taking further into account that practically all actual temperature measurements are performed by means of thermometers and not by statistical analysis of bodies, treated as statistical ensembles of elementary particles and excitations, an urgent need for a good phenomenological definition of temperature is thus evident.

It should be stressed here that the temperature is not a primary concept but, as was for the first time with sufficient plausibility shown by Mach [14], that the nearest experimentally accessible structure behind is so called ***hotness manifold*** ("Mannigfaltigkeit der Wärmezustände") the elements of which are ***hotness levels*** which are one-to-one related with experimentally observable ***thermoscopic states***. As was shown in our recent paper [15], the existence of such a set of thermoscopic states may be proved using an operational definition specifying the properties of a measuring device (thermoscope) together with the measurement conditions. The most salient among these conditions is that of ***thermal equilibrium*** [16] which is enabled by a ***thermal (diathermic) contact*** of the thermoscope with a body under investigation and the existence of which may be checked independently by so called ***correlation test*** [1]. It is an important consequence of the introduction of thermoscopes that we can experimentally observe the topological order of hotness levels as an order of quite another real physical quantity called ***thermoscopic variable*** (e.g. length of mercury thread, resistance, thermoelectric voltage). As these thermoscopic variables are real continuous quantities defined in certain closed intervals, this property is transferred by the said one-to-one relation also into the hotness manifold. The sewing-up of the different sets of thermoscopic states

---

[1] Correlation test is the procedure frequently used in practical thermometry which can be in general terms described as follows. Let us have in an inertial frame two systems, A and B, separated by a

macroscopically firm material partition defining their common boundary. Such a partition is called diathermic if the changes of system A induce the changes of system B and vice versa.

corresponding to the different thermoscopic variables, however, was shown not to be possible without exploitation of another physical entity, *fixed (thermometric) points*. The fixed point is a body prepared according to a definite prescription for which it is known that the thermoscope in diathermic contact with this body indicates reproducibly the well-defined thermoscopic state. The fixed points such as boiling point of helium, melting point of water, boiling point of water and melting point of platinum (all at normal atmospheric pressure) are examples well known from practical thermometry [17]. Because of the correspondence between the fixed points and some of thermoscopic states, the set of fixed points is ordered and, because fixed points are real bodies, also countable. Furthermore, an important empirical fact that the limitations put on the construction of the fixed points are not known [2] makes plausible the assumption that for each fixed point there is another "hotter" or another "colder" one and that there is an inter-lying fixed point between any two fixed points, sounds quite reasonably. Just such a property enables one not only to sew-up together the different overlapping parts of the hotness manifold corresponding to various thermoscopic variables but simultaneously provides a dense countable subset in the hotness manifold ensuring that this set is linear [18], and thus fully equivalent to the set of the real numbers (real axis).

Now we are ready to approach to the following definition: ***Temperature is any continuous one-to-one order-preserving mapping of hotness manifold on a simply connected subset of real numbers.*** It is evident from this definition that we are at enormous liberty to choose a particular temperature scale which thus rests entirely upon a convention. Only the hotness formally represented by hotness manifold has a right to be regarded as an entity existing in the Nature. (Cf.[14]) From this point of view our question, i.e. how the temperature, which is in fact a result of some arbitrary mapping, behaves by relativistic transformations, sounds as an ill-defined assignment. Indeed, the solution of the problem will depend on the particular way how the temperature is constructed and not only on the objective constraints. Moreover, it is quite clear that the rationality or irrationality of choice of the temperature scale will be decisive for further performance and intelligibility of theory of thermal effects. Therefore for the solution of our problem a detailed analysis of temperature concept in use is quite essential.

On the more-or-less historical and practical grounds, in other words, on the basis of fully arbitrary anthropomorphic criteria [15], a special mapping of the hotness manifold was chosen on an ordered subset of real numbers called the ***ideal (perfect) gas scale***. The equation controlling the behavior of the ideal gas, which is a hypothetical substance or concept rather than a real thing, reads:

---

[2] The temperatures observed range from $\sim 10^{-10}$ K (Low Temperature Lab, Helsinki University of Technology) up to $\sim 10^9$ K (supernova explosion) without any traces that the ultimate limits were actually reached. Speculative upper limit provides only the so called Planck temperature $T_P = \sqrt{(\hbar c/G)} \times (c^2/k) \approx 1.417 \times 10^{32}$ K, hypothetically corresponding to the first instant of Big Bang and depending on the assumption

that the constants involved are really universal. Therefore the conjecture presented in this paper, i.e. that the hotness manifold has no upper or lower bound, is obviously operating at least for all phenomena already known.

$$pV = n\mathrm{R}\,T,  \qquad (3)$$

where $p$ and $V$ are respectively the pressure and the volume of the ideal gas which alternatively play the role of thermoscopic variables. As the hypothetical thermoscope is considered a conventional gas thermometer [19] filled with $n$ moles, $n > 0$, of ideal gas. The constant R on the right side of equation (3), which need not be Lorentz invariant, has then a form of product R = k N where k and N are Boltzmann's and Avogadro's constants, respectively. The ideal gas temperature scale $T$ defined by means of equation (3) has some remarkable properties. For example, as both quantities $p$ and $V$ have a natural lower bound = 0, the temperature $T$ has also this lower bound and thus belongs to the class of **absolute temperatures** [20] for which one assumes that always $T > 0$. Notice that the possible value $T = 0$ is already excluded by our definition of temperature, because due to the absence of the lowest hotness level in the hotness manifold any continuous one-to-one order-preserving transformation has inevitably to map its improper point (i.e. $-\infty$) just on the point corresponding to absolute zero. **Nernst's law** of unattainability of absolute zero of temperature [21] is thus together with its consequences intrinsically involved in this definition of temperature.

## 3. Lorentz transformation of temperature

Having already at our disposal the prescription defining the temperature in mathematical terms, it is in principle possible to perform the Lorentz transformation of the left-hand side of equation (3) and to obtain in this way the formula for the relativistic transformation of temperature. The physical reasoning behind such a computation would not be, however, very transparent and convincing because the gas thermometer is rather a difficult device. Moreover, in order to be able to analyze also other temperature measuring methods (using e.g. platinum resistance, black-body radiation, thermoelectric voltage) in a sufficient generality the methodical approach is more relevant.

Fortunately, the properties of hotness manifold enable one to make the following fairly general considerations. First of all, it is evident that in order not to violate the Principle of Relativity the behavior of bodies realizing fixed thermometric points has to be the same in all inertial frames (Cf [22]). For example, it would be absurd to admit an idea that the water violently boiling in its rest system can simultaneously [3)] look calm if observed from another relatively moving inertial system. In other words, any fixed point has to correspond to the same hotness level regardless of the inertial frame used for the observation. Assigning, by means of some convention, to each body realizing the fixed point a certain "inventory entry", the resulting, by pure convention established list of numbers cannot be changed by a mere transfer from one inertial system to another.

[3] Notice that we have to do here with the essentially time-independent stationary process, where the Lorentz transformation of time plays no role. Let us also recall that the pressure, controlling e.g. boiling point of water, may be proved independently to be Lorentz invariant [23].

For example, using thus as an operational rule for stocktaking of fixed points formula (3) (in SI units with R = 8.3145 J/K mol) and assigning to the triple point of water an inventory entry 273.16 K, we obtain an ordered table of fiducial points of ideal gas scale (similar to the ITS [17]) which must be valid in all inertial frames. As the set of fixed points provides a dense subset (skeleton) in continuous hotness manifold, such a Lorentz-invariant table can be extended and detailed as we like and consequently, any hotness level can be, by means of this table, approximated with arbitrary accuracy. Due to the continuity of prescription (3) the whole ***ideal gas (Kelvin) scale T is then inevitably Lorentz invariant.***

The invariance of Kelvin scale has, however, a very interesting and far reaching consequence. Let us make the following "Gedankenexperiment" with two identically arranged gas thermometers both filled with one mole of ideal gas which are in two relatively moving inertial systems in diathermic contact with the same fixed point bath (for definiteness, with triple point of water) placed in their own frames. The pressure in both devices must be the same, because, as can be proved quite independently the pressure is Lorentz invariant [23]. Therefore we can write:

$$p = p_0, \qquad (4)$$

$$T = T_0, \qquad (5)$$

where index 0 is related, as above, to the quantities measured in the a-priori chosen rest system. Taking now the well-known Lorentz transformation of volume into account, we obtain from (3) the following series of equations

$$pV = p_0 \, V_0 \, \sqrt{(1 - v^2/c^2)} = RT = R_0 \, T_0 \, \sqrt{(1 - v^2/c^2)}, \qquad (6)$$

from which a somewhat astonishing relation immediately follows:

$$R = R_0\sqrt{(1 - v^2/c^2)}. \qquad (7)$$

The physical meaning of this formula is really far reaching. Taking into account, namely, that R is an entropy unit, equation (7) must simultaneously represent the transformation formula for entropy in general. This is, however, in severe contradiction with ***Planck's Ansatz*** claiming that the entropy is Lorentz invariant. We have to recall here that this Ansatz, serving as a starting point of numerous considerations in relativistic thermodynamics, has never been proved with sufficient exactness but from the beginning it was mere an intuitive conjecture [24]. (It was namely argued that the entropy has to be invariant, because it is the logarithm of a discrete number of states which is "naturally" Lorentz invariant.) Nevertheless, admitting once the relativistic invariance of temperature, we have to reject Planck's conjecture as unsound and particularly, we can also no more treat the various units of entropy, e.g. gas constant R and Boltzmann's constant k, as universal constants.

## 4. Distant measurement of temperature

The very task of the special theory of relativity is to study the transformation laws connecting the experimental results of observers in different inertial systems performing the same measuring operations. Frequent types of such measurements are so called "***distant measurements***" the aim of which is to determine the physical quantity belonging to a certain moving inertial system by means of measurements made at the rest system. The operational methods for distant measurement of e.g. length, time, and intensity of fields are generally known. In the case of temperature, however, due to its special physical nature we encounter some peculiar difficulties. The main problem, which is intuitively not quite obvious, is the principal impossibility to establish the thermal equilibrium between two relatively moving inertial systems. Namely, the relative movement of systems A and B (see footnote 1) prevents one from answering without ambiguity, on the basis of correlation test, the question whether the common boundary is diathermic or not, which makes any judgment on the thermal equilibrium quite questionable. Indeed, it is clear that the boundary between two relatively moving systems has to move at least with respect to one of them. In such a case, however, the interaction between these systems can exist even if the boundary is non-diathermic (adiabatic). For example, the moving boundary can exert a pressure on one of the systems without changing the state of the other and/or a charged system A surrounded by a metallic envelope, regardless of the fact whether it is diathermic or adiabatic, can induce dissipative equalization currents in system B without affecting the charge distribution inside system A. In order to exclude such cases, the temperature of any body must be measured only by means of a thermometer which is in the rest with respect to the body, and this operation cannot be, in principle, performed by a relatively moving observer [4] (cf. also [25]). Hence the temperature cannot be the subject of a direct distant measurement in principle. It can only be the result of local measurement and subsequent data transfer into another inertial system. (If possible, the digital mailing of the data would be the best choice.) Of course, as the theoretical basis for the determination of temperature of moving objects equation (5), expressing the Lorentz invariance of temperature, has to be simultaneously taken into account. The operational rules for distant measurement of temperature may then be formulated as follows:

*1) Bring the measured body in diathermic contact with the thermometer in their common inertial rest frame.*
*2) Reconstruct in the relatively moving inertial system the reading of the thermometer applying transformation rules relevant to the thermoscopic variable used.*

Obviously, such a two-step procedure should ensure the consistent results even if different thermometers are used. If properly chosen and correctly transformed, namely, the thermoscopic variable must reproduce the same thermometer reading in any inertial system.

[4]) Interestingly enough, the similar assumption that the temperature measurement is possible just only if the thermometer is in rest with respect to the measured system is as self-evident, without any proof, used in recent theoretical literature, see e.g. [26].

In order to illustrate the application of the above formulated rules and especially to exemplify the importance of proper choice of thermoscopic variable, the temperature measurement by means of optical pyrometry has been chosen. This example is very instructive because it clearly shows, beside the general features of this technique, some of its deceptive aspects as well. Moreover, it is also closely related to the problem of Mosengeil-Ott's antinomy mentioned in the title.

Optical pyrometry is in its simplest form based on the so called **Wien's displacement law** which may be written in terms of frequency as:

$$T_0 = \hbar \omega_{M0} / 2.82 \, k, \qquad (8)$$

where $\omega_{M0}$ is the frequency corresponding to the maximum of equilibrium distribution of black-body radiation [27]. As a thermometer serves the cavity with a small opening attached to the measured body, or alternatively, simply the surface of the body itself which is assumed to be "black". The emitted light is then in the rest system of the body analyzed by means of a spectrometer and the frequency $\omega_{M0}$ corresponding to the maximal radiation power is find out. After that the temperature $T_0$ of the body can be immediately determined using equation (8). If we investigate the radiation of black-body thermometer, being observers in another relatively moving inertial system, the thermoscopic variable $\omega_M$ should be, according to our rule 2), transformed into original $\omega_{M0}$. It is a well known fact that the relativistic transformation of frequency is reduced to the multiplication by Doppler's factor $K(v,\theta)$ i.e.

$$\omega_{M0} = \omega_M \, K(v,\theta), \qquad (9)$$

where $v$ is the relative velocity of the motion and $\theta$ the angle between axis of motion and the direction of observation [28]. Writing then Doppler's factor in *scriptio plena* and substituting the resulting $\omega_{M0}$ into relation (8), we obtain a formula which is normally used for the analysis of *relict radiation* [29], namely

$$T_0 = T \{1- (v/c) \cos \theta \} / \sqrt{(1 – v^2/c^2)}. \qquad (10)$$

It is apparent at first glance that for measurement performed in the direction perpendicular to the relative velocity vector (i.e. for $\theta = \pi/2$) formula (10) becomes identical with Ott's relation (2). How can be, however, this result reconciled with our assertion that the temperature is Lorentz invariant? We claim that this discrepancy is due to the improper choice of thermoscopic variable. We have, namely, tacitly made an incorrect assumption that the shape of distribution of black-body radiation is Lorentz invariant. As was, however, convincingly shown e.g. by Boyer [30] the only Lorentz invariant part of Planck's distribution is that represented by the so called zero-point temperature independent term $(\hbar/2\pi^2 c^3)\omega^3 d\omega$. Just in contrast, the temperature dependent term of black-body radiation distribution observed from a moving inertial system is skewed a little bit loosing thus the affinity to Planck's function. Consequently, the frequencies corresponding to the maxima of radiation in the rest and moving frames are no more connected by means of equation (9) and cannot thus serve as a "good" thermoscopic variable. Instead, in order to determine parameters of Planck's distribution belonging to the rest system (the temperature $T_0$ involved) it must be reconstructed from the complete distribution observed in the moving system. In other words, in the case of optical pyrometry the role of the thermoscopic variable cannot play a single point but the distribution as a whole.

## 5. Conclusions

Analyzing anew the concept of phenomenological temperature we have been able to demonstrate that the Mosengeil-Ott's antinomy is actually an artefact, while the Kelvin temperature has to be Lorentz invariant. It has been further shown that in such a case the extensive variable conjugate to the temperature, i.e. entropy, can no more be Lorentz invariant (violation of Planck's Ansatz). Besides, operational rules for distant measurement of temperature were formulated and applied to the practically important case of optical pyrometry.

## Acknowledgement

This work was supported by the Grant Agency of ASCR Contract No IAA100100639 and by IP-ASCR v. v. i., in the frame of Institutional Research Plan No AV0Z10100521.